\documentstyle[preprint,aps]{revtex}

\begin{document}




\title{Effective nonlinear model of 
resonant tunneling nanostructures}

\author{Enrique Diez and Angel S\'{a}nchez}

\address{Departamento de Matem\'aticas, Universidad
Carlos III de Madrid, C./ Butarque 15, 
E-28911 Legan\'{e}s,
Madrid, Spain}

\author{Francisco Dom\'{\i}nguez-Adame}

\address{Departamento de F\'{\i}sica de Materiales,
Facultad de F\'{\i}sicas, Universidad Complutense, 
E-28040 Madrid, Spain}

\maketitle

\begin{abstract}

We introduce a model of a nonlinear double-barrier
structure, to describe in a simple way the effects of 
electron-electron scattering while remaining analytically tractable.
The model is based on
a generalized effective-mass equation where a  nonlinear local field interaction is
introduced to account for those inelastic scattering phenomena. 
Resonance peaks seen in the transmission coefficient spectra for the linear case appear 
shifted to higher energies depending on the magnitude of the nonlinear coupling.
Our results are in good agreement with self-consistent solutions of the
Schr\"odinger and Poisson equations. The calculation procedure is seen to
be very fast, which makes our technique a good candidate for rapid
approximate analysis of these structures.

\end{abstract}


\narrowtext


Resonant tunneling (RT) through double-barrier structures (DBS) make
these systems very promising candidates for a new generation of
ultra-high speed electronic devices.  For instance, a
GaAs-Ga$_{1-x}$Al$_x$As DBS operating at THz frequencies has been
reported in the literature.\cite{Sollner} The basic reason for RT to
arise in DBS is a quantum phenomenon whose fundamental characteristics
are by now well understood: There exists a dramatic increase of the
electron transmittivity whenever the energy of the incident electron is
close to one of the unoccupied quasi-bound-states inside the
well.\cite{Ricco} In practice, a bias voltage is applied to shift the
energy of this quasi-bound-state of nonzero width so that its center
matches the Fermi level.  Consequently, the $j-V$ characteristics
present negative differential resistance (NDR).

In actual samples, however, the situation is much more complex than this
simple picture.  This is so even in good-quality heterostructures, when
scattering by dislocations or surface roughness is negligible.  In
particular, {\em inelastic} scattering is always present in real
devices.  An example of inelastic scattering events is
electron-electron interaction, in which the energy of the tunneling
electron changes and the phase memory is lost.  The influence of such
many-body effects on DBS has recently attracted considerable attention.
Even with its rather satisfactory degree of success, many-body
calculations have difficulties that, in some cases, may complicate the
interpretation of the underlying physical processes.
Recently Presilla {\em et al.}\cite{Presilla} 
pointed out the possibility of nonlocal effective nonlinearities
due to many-body interactions in electron transport through DBS.
Several results have been obtained using this mean field analysis such as quantum
chaos\cite{Jona} and  nonlinear oscillations.\cite{Jona2,Sun} 
Loosely speaking, this kind of treatment could
be regarded as similar to Hartree-Fock and other self-consistent
techniques, which substitute many-body interactions by a nonlinear
effective potential. It has to be stressed, however, that
the nonlocal interaction might not be the most suitable one in
many contexts (e.g., when wells are wide) because it does not take
into account the spatial variation of the effective potential.
In any event what appears to be clear is that RT in nanostructures
is intrinsically a (perhaps weakly) non-linear phenomena  as have been 
shown in many different situations, such as
the nonlinearity dependence of the lock-in voltage with the number of
tunneling electrons obtained with a Hartree-Fock approximation by Wang {\em 
et al.}\cite{Bishop}

In this letter we present a model where the nonlinear interaction is driven by a local field
instead of the
mean-field approach used by Presilla and co-workers.\cite{Presilla}
To describe our model we have chosen a 
DBS under an applied electric field.  The
thickness of the whole structure is $L$ and the thickness of the well is
$d$.  The barriers are assumed to be of the same thickness (symmetric
case) but as will be evident below this is not a restriction of our
approach.  The structure is embedded in a highly doped material acting
as contact, so that the electric field is applied only in the DBS. We
focus on electron states close to the bandgap and thus we can neglect
nonparabolicity effects hereafter.  Then the one-band effective-mass
framework is completely justified to get accurate results.  For the sake
of simplicity, we will further assume that the electron effective-mass
$m^*$ is constant through the whole structure.  This hypothesis is
related to the fact that for the time being,
we are not interested in high quantitative
accuracy, although we note that the spatial dependence
of the effective mass can
be taken into account if necessary.

The generalized 
effective-mass equation is given by [we use units such that 
energies are measured in
effective Rydberg (Ry$^*$) and lengths in effective Bohr radius (a$^*$),
being $1\,$Ry$^*=5.5\,$meV and $1\,$a$^*=100\,$\AA\ in GaAs] 

\begin{equation}
-\psi_{zz}(z)+\left[V_{Eff}(z) - E\right]\,\psi(z)=0,
\label{Sch}
\end{equation}
where $V_{Eff}(z)$ is the potential term which we discuss below.
The DBS can be regarded as an effective
medium which reacts to the presence of the tunneling electron, leading
to a feedback mechanism by which inelastic scattering processes change
the RT characteristics of the device.  It thus follows that $V_{Eff}(z)$ must
contain nonlinear terms if it is to summarize the medium reaction which
comes from the electron-electron and electron-lattice interactions. This
term will be a generalized functional of the electronic probability density
$\cal{F}$$(|\psi(z)|^2)$. We can now expand this functional as power
series in 
$|\psi(z)|^2$, and neglecting higher order  (which implies an assumption of
weak nonlinearity or weak electron-electron scattering) contributions
we postulate that the potential in Eq.~(\ref{Sch}) has the form
\begin{equation}
\label{Veff}
V_{Eff}(z) = V_0\left[\chi(z) + \tilde{\alpha}|\psi(z)|^2\right] - eFz,
\end{equation}
where $V_0$ is the conduction band-offset at the interfaces, 
$F$ is the
electric field applied along the growth direction, and
$\chi(z)$ is the characteristic
function of the barriers,
\begin{equation}
\chi(z)=\left\{\begin{array}{ll} 1, & \mbox{\rm if}\ 0<z<(L-d)/2,\\
                                   1, & \mbox{\rm if}\ (L+d)/2<z<L,\\
                                   0, & \mbox{otherwise},
                 \end{array} \right.
\label{charab}
\end{equation}
and all the nonlinear physics is contained in the coefficient
$\tilde{\alpha}$ which we discuss below.  

There are two factors that configure the medium nonlinear response to the
tunneling electron.  First, it goes without saying that there are
repulsive electron-electron Coulomb interactions, which should enter the
effective potential with a positive nonlinearity, 
i.e., the energy is increased by local charge accumulations, leading to
a positive sign for $\tilde{\alpha}$.
Therefore intuitively, a negative sign for $\tilde{\alpha}$ does not appear
to be realistic.
Nevertheless, we consider also that situation because in some materials,
like polar semiconductors, the electron polarizes the surrounding medium
creating a local, positive charge density. Hence the electron reacts to
this polarization and experiences an attractive potential.
This
happens, for instance, in the polaron problem in the weak coupling
limit, which becomes valid in most semiconductors, and where it can be
seen that the lowest band energy state decreases. \cite{Callaway}


We now work starting from Eq.~(\ref{Sch}) with the definition in
Eq.~(\ref{Veff}) to cast our equations in a more tractable form.  For
simplicity, and because we are interested in intrinsic DBS features, we
consider that the contacts in which the structure is embedded behave
linearly.  Therefore, the solution of Eq.~(\ref{Sch}) is a linear
combination of traveling waves.  As usual in scattering problems, we
assume an electron incident from the left and define the reflection,
$r$, and transmission, $t$, amplitudes by the relationships
\begin{equation}
\psi(z)=\left\{ \begin{array}{ll} A\left(e^{ik_0z}+re^{-ik_0z}\right)
       & z<0, \\ Ate^{ik_Lz} & z>L,  \end{array} \right.
\label{solution}
\end{equation}
where $k_0^2=E$, $k_L^2=E+eFL$, and $A$ is the incident wave amplitude.
Now we define $\psi(z)=(A|t|\sqrt{k_L})\phi(z)$, 
$\alpha=\tilde{\alpha}\, k_L \, |A t|^2$. 
Notice that
$\alpha$ depends on the amplitudes of the incoming and
outgoing waves, which will be relevant later. 
Using Eq.~(\ref{Sch}) we get
\begin{equation}
-\phi_{zz}(z)+[V_{Eff}(z)-E]\,\phi(z)=0,
\label{fi}
\end{equation}
where now (\ref{Veff}) read as
\begin{equation}
\label{Veff2}
V_{Eff}(z) = V_0\left[\chi(z) + \alpha|\phi(z)|^2\right] - eFz.
\end{equation}
To solve the scattering problem in the DBS  we develop a
similar approach to that given in Ref.~\onlinecite{Knapp}.  Since
$\phi(z)$ is a complex function, we write $\phi(z) = q(z) \exp[
i\gamma(z)]$, where $q(z)$ and $\gamma(z)$ are real functions.
Inserting this factorization in Eq.~(\ref{fi}) we have $\gamma_z (z) =
q^{-2}(z)$ and
\begin{eqnarray}
-q_{zz}(z)+{1\over q^3(z)}+\left[V_0\chi(z)-eFz-E\right]\,q(z) & + &
\nonumber \\
+V_0 \alpha q^3(z) & = & 0.
\label{q}
\end{eqnarray}
This nonlinear differential equation must be supplemented by appropriate
boundary conditions. However, using Eq.~(\ref{solution}) this problem
can be converted into a initial conditions equation. In fact, it is
straightforward to prove that
\begin{equation}
\label{ic}
q(L)=k_L^{-1/2},\>q_z(L)=0,
\end{equation}
and that the transmission coefficient is given by
\begin{equation}
\tau=\,{4k_0q^2(0)\over 1+2k_0q^2(0)+k_0^2
q^4(0)+q^2(0)q_z^2(0)}.
\label{tau}
\end{equation}
Hence, we can integrate numerically (\ref{q}) with initial conditions
(\ref{ic}) backwards, from $z=L$ up to $z=0$, to obtain $q(0)$ and
$q_z(0)$, thus computing the transmission coefficient for given
nonlinear coupling $\alpha$, incoming energy $E$ and
applied voltage $V=FL$.

Once the transmission coefficient has been computed, and recalling that
contacts are linear media, the tunneling current density at a given
temperature $T$ for the DBS can be calculated within the
stationary-state model from
\begin{mathletters}
\label{eq2}
\begin{equation}
j(V)={m^*ek_BT\over 2\pi^2\hbar^3}\,\int_0^\infty\> \tau(E,V)N(E,V)\,dE,
\label{eq2a}
\end{equation}
where $N(E,V)$ accounts for the occupation of states to both sides of
the device, according to the Fermi distribution function, and it is
given by
\begin{equation}
N(E,V)=\ln\left(\frac{1+\exp[(E_F-E)/k_BT]}{1+\exp[(E_F-E-eV)/k_BT]}
\right),
\label{eq2b}
\end{equation}
\end{mathletters}
where $k_B$ is the Boltzmann constant.


In our calculations we have considered a GaAs-Ga$_{0.65}$Al$_{0.35}$As
double-barrier structure with $L=3d=150\,$\AA.  The conduction-band
offset is $V_0=250\,$meV.  In the absence of applied electric field and
nonlinearities, there exists a single, very narrow resonance with
$\tau\sim 1$ below the top of the barrier, with an energy of
$\sim 81\,$meV, and hence the well supports a single quasi-bound state.
Figure 1 shows the transmission coefficient as a function of the incoming
energy for different values of the nonlinear coupling $\alpha$ (a) $0$ and $10^{-4}$, (b) $10^{-3}$
and (c) $10^{-2}$, at zero bias. Insets show
the effective potential $V_{Eff}$ at the energy marked with an arrow.
It is clear that the resonances are shifted to energies higher than in the noninteracting
case. The shift is produced by the accumulation of charge in the well as shown in the figures
of $V_{Eff}$. These results are in very good agreement with 
self-consistent calculations.\cite{cota,Mark}
Note that $V_{Eff}$ reproduce the charge accumulation
in the barriers, close to the heterojunctions, and in the center of
the well, obtained in the Hartree approximation.\cite{Bastard}
For completeness, we consider also the case when 
lattice lattice polarization
effects are stronger than electron-electron interactions, thus leading to a
negative $\alpha$. This regime may give rise to some kind of unphysical 
``superconducting instability'' 
but for small values of the nonlinear coefficients we have  never found 
that problem.
Figure 2 shows the transmission coefficient as a function of the incoming
energy for different values of the nonlinear coupling $\alpha$ (a) $0$ and -$10^{-4}$, (b) -$10^{-3}$
and (c) -$10^{-2}$, at zero bias. Now the resonances are shifted to lower energies.
Also we show that for some values of the nonlinear coefficient there appears another
quasi-bound state for lower energies. Looking at the effective potential we 
conclude that the polarization
of the lattice produce a deeper well and consequently the possibility of
new bound states, thus shifting or even doubling the original resonance.

To apply our calculations to obtain measurable quantities, we now discuss
the $j-V$ characteristics, as computed from (\ref{eq2}).
When voltage is applied, the energy of the quasi-bound state level is
lowered and a strong enhancement of the current arises whenever the
Fermi level matches this resonance, thus leading to the well-known RT
phenomenon.
Typical results are shown in Fig.~\ref{jv} for the particular values 
(a) $\alpha=10^{-3}$ and (b) $-10^{-3}$. 
The temperature was 77 K and Fermi energy  was (a) $E_F$ = $83$ meV and (b) $3$ meV.
In both cases we see clear NDR signatures. For the repulsive interaction
we have a NDR peak shifted to higher energies (note the large Fermi energy) 
than in the linear case as  
expected in view of Fig~\ref{tra1}. The attractive self-interaction causes
the occurrence of a second peak at a lower voltage clearly related to the sideband
in the transmission coefficient [Fig ~\ref{tra2}(b)].
A study of bistability\cite{Sun} may be carried out whitin our model;
this kind of results will be obtained in a complete dynamical study of
the problem that is in progress.

We have demonstrated that the effective nonlinear interaction we have
introduced captures the
essential physics of some inelastic effects in RT in a very simple way.
The virtue of such an approach is that it allows to gain insight on the
features of DBS without the burden of intensive computations providing,
in an inexpensive way, a qualitative picture of what is to be expected in particular
devices. Our model introduces a local field interaction that 
reproduce the effective potential
in a more realistic way than mean-field approximation proposed 
in Ref.~\onlinecite{Presilla}.
Our model can be easily modified to study other heterostructures in an electric field,
like superlattices, and including different kinds of impurities and disorder.
An important conclusion we draw from this work is that comparison
of the result of current measurements with theoretical expectations
allows to conclude which effect or group effects (electron-electron interaction
or lattice polarization) is responsible for deviations from linearity
in any specific material, by just looking at the shifting of the
peaks or the appearance of new peaks. Also, it has to be stressed that
a mere quantitative comparison between the present work and our idea of
effective potentials to Hartree results on the same structure is needed,
and we plan to elaborate further on this point in the new future.
In addition to realize the study of the interplay between disorder and nonlinearity in this kind
of models will be very interesting because of the possibility of 
finding delocalization, driven
by nonlinear effects.\cite{todos} 


We are very thankful to Paco Padilla and Alan R. Bishop for useful discussions. 
This work is supported from CICYT (Spain) through project No.~MAT95-0325.


\begin{figure}
\caption{Transmission coefficient $\tau$ as a function of the
electron energy at zero bias for (a) $\alpha=10^{-4}$, (b) $10^{-3}$, and (c) $1
0^{-2}$.
For comparison, dashed line indicates in (a) the result for $\alpha=0$.
Insets show
the effective potential $V_{Eff}$ at the energy marked with an arrow.}
\label{tra1}
\end{figure}

\begin{figure}
\caption{Transmission coefficient $\tau$ as a function of the
electron energy at zero bias for (a) $\alpha=-10^{-4}$, (b) $-10^{-3}$, and (c)
$-10^{-2}$.
For comparison, dashed line indicates in (a) the result for $\alpha=0$.
Insets show
the effective potential $V_{Eff}$ at the energy marked with an arrow. Left inset
in (c)
shows an enlarged view of the lower resonant peak.}
\label{tra2}
\end{figure}

\begin{figure}
\caption{ Computed $j-V$ characteristics for $T = 77 K$, 
with (a) $\alpha = 10^{-3}$,$E_F$ = 83 meV  and (b) 
$\alpha = -10^{-3}$, $E_F$ = 3 meV.}
\label{jv}
\end{figure}


\begin{references}

\bibitem{Sollner} T.\ C.\ L.\ G.\ Sollner, W.\ D.\ Goodhue, P.\ E.\
Tannenwald, C.\ D.\ Parker, and D.\ D.\ Peck, Appl.\ Phys.\ Lett.\ {\bf
43}, 588 (1984).

\bibitem{Ricco} B.\ Ricco and M.\ Ya.\ Azbel, Phys.\ Rev.\ B {\bf 29},
1970 (1984).

\bibitem{Presilla} C.\ Presilla, G.\ Jona-Lasinio, and F.\ Capasso,
Phys.\ Rev.\ B {\bf 43}, 5200 (1991)

\bibitem{Jona} G.\ Jona-Lasinio, C.\ Presilla,
and F.\ Capasso, Phys.\ Rev.\ Lett.\ {\bf 68}, 2269 (1992).

\bibitem{Jona2} G.\ Jona-Lasinio, C.\ Presilla, and J.\ Sj\"ostrand,
Annals of Physics {\bf 240}, 1 (1995).

\bibitem{Sun} N.\ G.\ Sun and G.\ P.\ Tsironis, Phys.\ Rev.\ B {\bf 51},
11\,221 (1995).

\bibitem{Bishop} L.\ Wang, J.\ K.\ Zhang and A.\ R.\ Bishop,
Phys.\ Rev.\ Lett.\ {\bf 74}, 4710 (1995).

\bibitem{Callaway} J.\ Callaway, {\em Quantum Theory of the Solid
State} (Academic Press, CA, 1991), p 711.

\bibitem{Knapp} R.\ Knapp, G.\ Papanicolaou, and B.\ White, J.\ Stat.\
Phys.\ {\bf 63}, 567 (1991).

\bibitem{cota} E.\ Cota and S.\ E.\ Ulloa, 
Phys. Rev. B {\bf 51}, 10\,875 (1995).

\bibitem{Mark} M.\ S.\ Sherwin, K.\ Craig, B.\ Galdrikian, J.\ Heyman,
A.\ Markelz, K.\ Campman, S.\ Fafard, P.\ F.\ Hopkins, and A.\ Gossard,
Physica D {\bf 83}, 229 (1995).

\bibitem{Bastard} G.\ Bastard, {\em Wave mechanics applied to semiconductor
heterostructures} (Les Editions de Physique, France, 1988), p 161.

\bibitem{todos} A.\ S\'anchez and L.\ V\'azquez, Int.\ J.\ Mod.\
Phys.\ B {\bf 5}, 2825 (1991); S.\ A.\ Gredeskul and Yu.\ S.\ Kivshar,
Phys.\ Rep.\ {\bf 216}, 1 (1992); E.\ Diez, F.\ Dom\'\i nguez-Adame, and
A.\ S\'an\-chez, Phys.\ Lett.\ A {\bf 198}, 403 (1995).

\end{references}
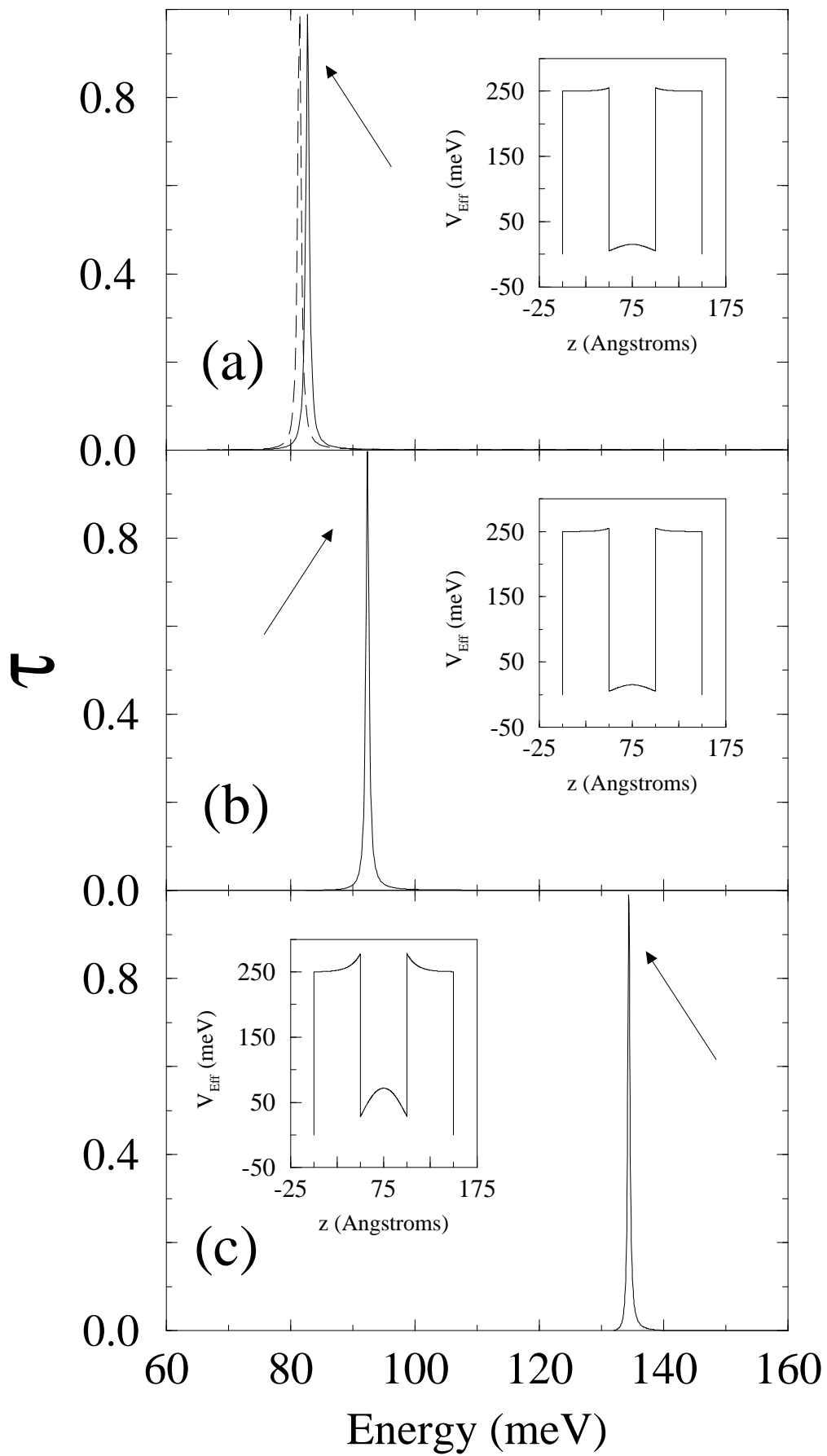
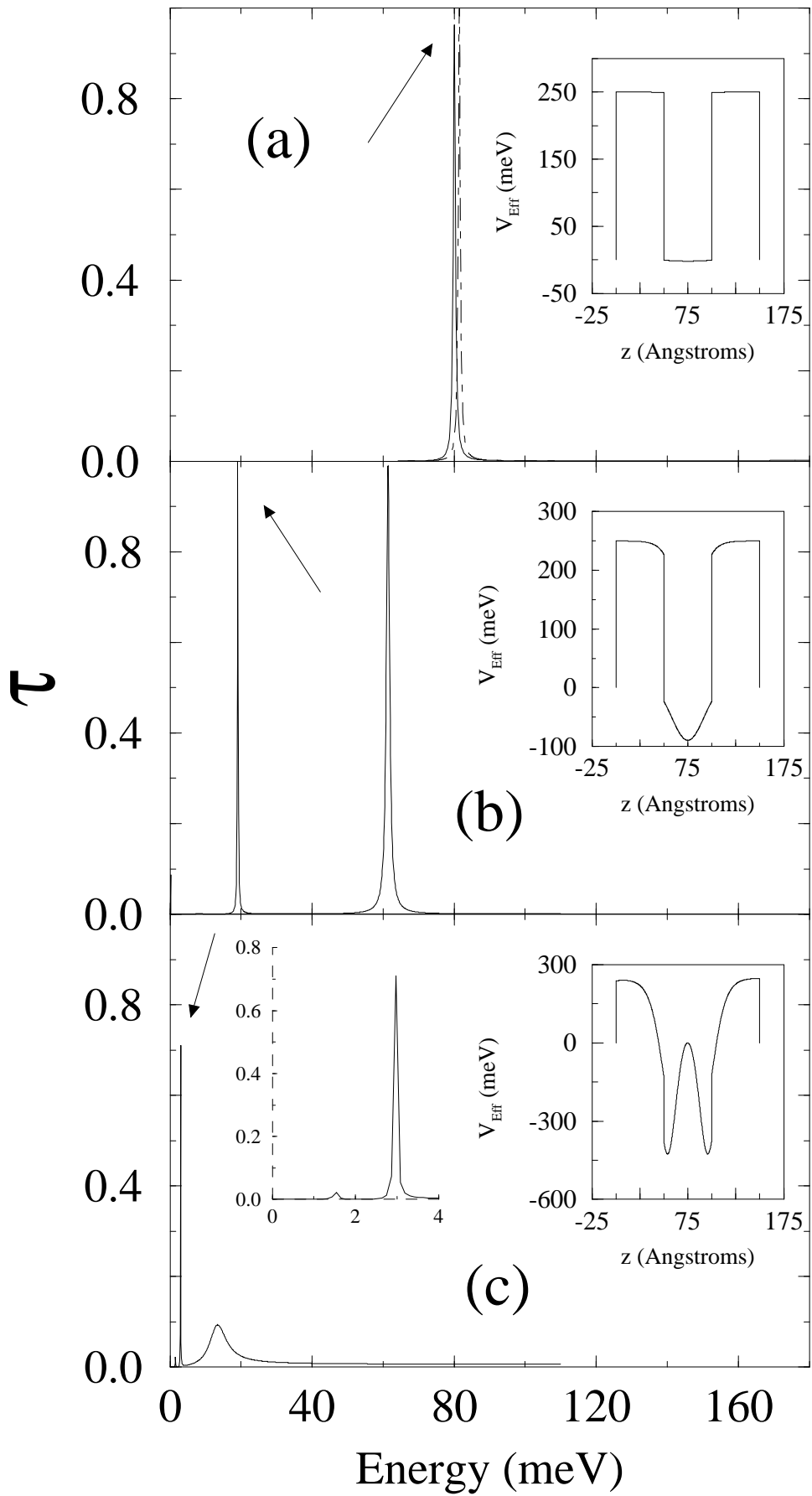
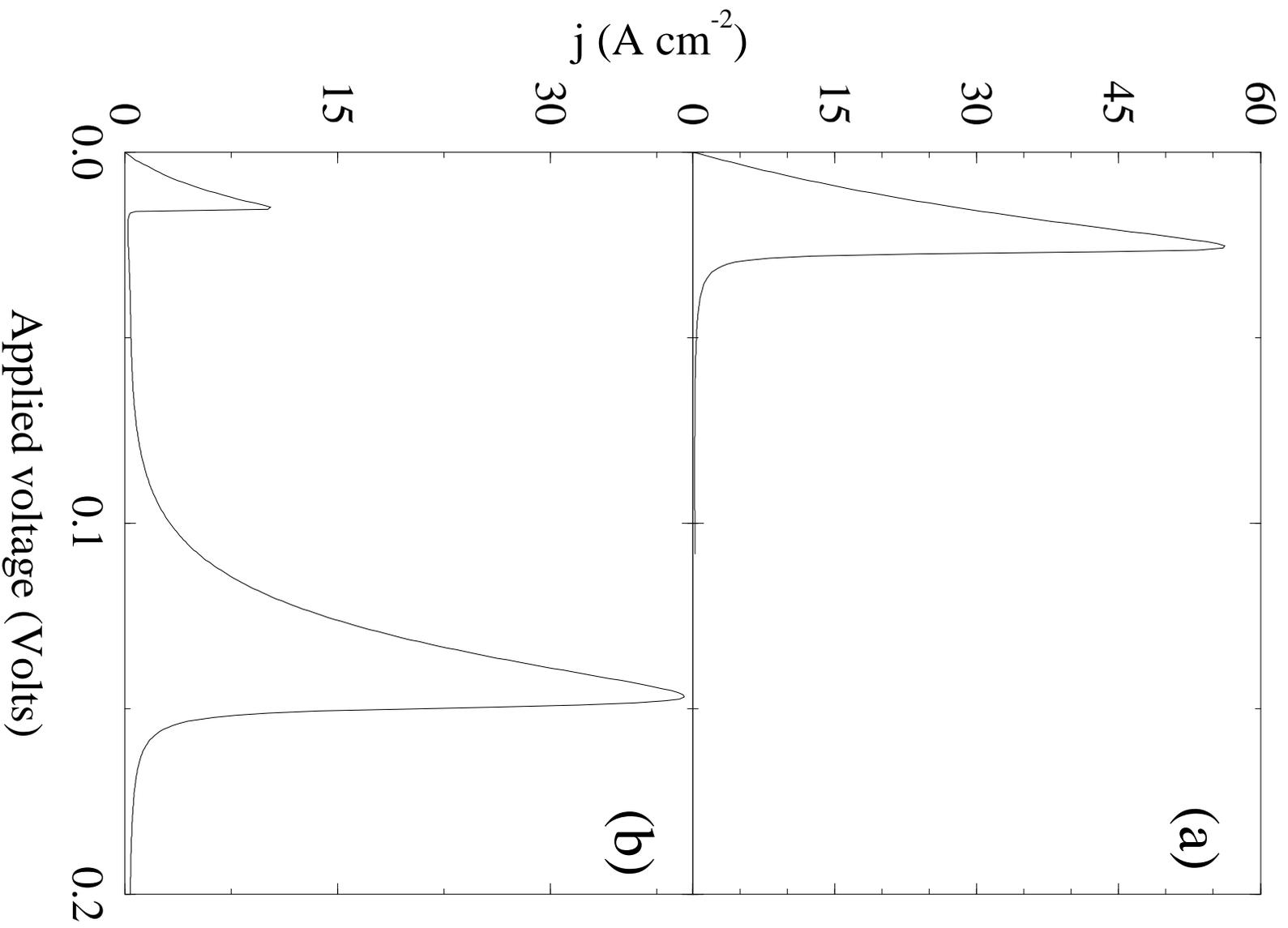
\end{document}